\documentclass[aps,prl,showpacs,floatfix,twocolumn,superscriptaddress]{revtex4}
\usepackage{graphicx}
\usepackage{dcolumn}
\usepackage[mathscr]{eucal}
\usepackage{amsmath,amssymb,bbm}
\usepackage[latin1]{inputenc}
\usepackage[normalem]{ulem}

\begin{document}
\title{Photonic Hall effect in cold atomic clouds}

\author{Beno\^it Gr\'emaud}
\affiliation{Laboratoire Kastler-Brossel, Universit\'e Pierre et Marie Curie-Paris 6, ENS, CNRS;
4 Place Jussieu, F-75005 Paris, France}
\affiliation{IPAL, CNRS, I2R;  1 Fusionopolis Way, Singapore 138632}
\author{Dominique Delande}
\affiliation{Laboratoire Kastler-Brossel, Universit\'e Pierre et Marie Curie-Paris 6, ENS, CNRS;
4 Place Jussieu, F-75005 Paris, France}
\author{Olivier Sigwarth}
\affiliation{Laboratoire Kastler-Brossel, Universit\'e Pierre et Marie Curie-Paris 6, ENS, CNRS;
4 Place Jussieu, F-75005 Paris, France}
\author{Christian Miniatura}
\affiliation{IPAL, CNRS,  I2R; 1 Fusionopolis Way, Singapore 138632}
\affiliation{Institut Non Lin\'eaire de Nice, UMR 6618, Universit\'e de Nice Sophia, CNRS; 1361 route
des Lucioles, F-06560 Valbonne}

\date{\today}
 
\begin{abstract}
  \textbf{Abstract}: On the basis of exact numerical simulations and
  analytical calculations, we describe qualitatively and
  quantitatively the interference processes at the origin of the
  photonic Hall effect for resonant Rayleigh (point-dipole) scatterers in a
  magnetic field.  For resonant incoming light, the induced giant 
  magneto-optical effects result in relative Hall currents in the percent 
  range, three orders of magnitude larger than with classical scatterers. This suggests
  that the observation of the photonic Hall effect in cold atomic vapors is within experimental reach.
\end{abstract}

\pacs{42.25.Dd, 32.60.+i, 78.20.Ls,32.10.Dk}

\maketitle

Light propagation in homogeneous media in the presence of a static magnetic
field is a rich and vivid field of research where the symmetries dictated by
nature lead to subtle magneto-optical phenomena \cite{landau}. About ten years
ago, the question of magneto-optics in strongly scattering media was addressed
and several effects bearing close analogies with electronic transport were
theoretically predicted and observed \cite{tiglari}. One striking example is
given by the so-called photonic Hall effect (PHE) where light propagating in a
scattering medium subject to a transverse magnetic field can be deflected in the
direction perpendicular to both the incident beam and the magnetic field
\cite{prl75vT, nature381RT}. Cold atomic gases provide an appealing testing
ground for such effects in the multiple scattering regime. Indeed they
constitute a perfect monodisperse sample of highly-resonant point scatterers
which are very sensitive to external perturbations and where spurious
phase-breaking mechanisms can be easily circumvented. Typically few Gauss are
enough to induce strong magneto-optical effects like the Faraday rotation
\cite{faraday} in sharp contrast with classical materials where Teslas are
needed to induce significant effects. If the impact of a magnetic field on
coherent backscattering has already been studied~\cite{hanle,cbsB}, the
question of the observation of the PHE in atomic vapors is still open. In this
paper, we present analytical and numerical calculations identifying the physical
origin of the PHE for point-dipole scatterers without internal degeneracies. Our
results show that the effect should be observable in cold atomic gases. 

For a quantitative study of the PHE, one needs to address the question of
directional asymmetries displayed by the configuration-averaged radiation
pattern of an assembly of atoms located at random positions and illuminated by
an incident monochromatic plane wave  (wave vector $\mathbf{k}$, polarization
vector $ \boldsymbol{\epsilon} \perp \mathbf{k}$, angular frequency $\omega = c
k=2\pi c/\lambda$) while being subjected to an external static magnetic field
with strength $B$ pointing in the direction $\hat{\mathbf{B}}$. We consider here
the simplest possible atomic internal structure serving our purposes, namely
two-level atoms having a groundstate with angular momentum $J=0$ 
connected by a narrow optical dipole transition to an excited state with angular
momentum $J_e =1$. The energy separation between the atomic states is $\hbar
\omega_0 = h c/\lambda_0$ and the natural energy width of the excited state is
$\hbar \Gamma \ll \hbar \omega_0$. This is one of the best possible natural
realizations of resonant point scatterers~\cite{pre53TMN} and it corresponds for
example to the case of ${}^{88}\mathrm{Sr}$ atoms ($\lambda_0=461$nm,
$\Gamma/2\pi=32$MHz, Land\'e factor of the excited state $g_e=1$). When $B=0$,
the incident light is quasi-resonant ($\lambda \approx \lambda_0$) with this
optical dipole transition and we will denote by $\delta = (\omega - \omega_0)$
the light detuning with respect to the atomic line ($\delta \ll \omega_0$). In
the
rest of the paper, we assume that the incident light intensity is low
enough to neglect all nonlinear effects.  When the magnetic field is applied,
the internal degeneracy is lifted (Zeeman effect) and the excited level is split
into 3 components separated by $\mu B$ where $\mu/2\pi =1.4$ MHz/G is the Zeeman
shift rate. As soon as the Zeeman shift becomes comparable to the resonance
width, i.e  $\phi_B = 2\mu B/\Gamma \sim 1$, the scattering properties of each atom
are strongly modified (this would occur at $B \sim 11$~G in the case of
${}^{88}\mathrm{Sr}$).

The source of the field radiated by the atom is the oscillating electric dipole
moment $\mathbf{d} \, \exp(-i\omega t)$ induced by the incident electric field
$\mathbf{E} \exp(-i\omega t)$. The radiated spectrum is elastic
because there is no Zeeman effect in the groundstate. The situation would be
 more involved for atoms with a degenerate groundstate where
frequency changes are possible, leading to an inelastic spectrum. In our
situation, $\mathbf{d} = \epsilon_0 \,
\underline{\boldsymbol{\alpha}}(\mathbf{B})\mathbf{E}$ and the radiation
properties are fully characterized by the polarizability tensor
$\underline{\boldsymbol{\alpha}}(\mathbf{B})= \alpha_0 \,
\mathcal{T}(\mathbf{B})$ given by:
\begin{equation}
\label{polartensor}
\begin{aligned}
&\alpha_0 = \frac{6\pi}{k^3} \, \frac{\Gamma/2}{\delta+i\Gamma/2}\\
&\mathcal{T}(\mathbf{B}) = \zeta(B) \, \mathbbm{1} + \eta(B) \, \mathbbm{1}
\times \hat{\mathbf{B}} + \xi(B) \, \hat{\mathbf{B}}\hat{\mathbf{B}}
\end{aligned}
\end{equation}
where $\alpha_0$ is the complex atomic polarizability at $B=0$. The dyadic
tensor $\mathcal{T}(\mathbf{B})$ embodies the effect of the magnetic field on
the photon polarization degrees of freedom and gives rise to the usual
magneto-optical effects. The $\zeta$ term is responsible for the normal
extinction of the forward beam in a scattering medium made of such atoms
(Lambert-Beer law). The $\eta$ term describes the magnetically-induced rotation
of the atomic dipole moment (Hanle effect) \cite{hanle} and induces Faraday
rotation and dichroism effects in the forward beam when $\mathbf{k} \parallel
\mathbf{B}$ \cite{landau}. Finally, the $\xi$ term  is responsible for the
Cotton-Mouton effect also observed in the forward beam when $\mathbf{k} \perp
\mathbf{B}$ \cite{landau}. These coefficients read:
\begin{equation}
\zeta \! = \! \frac{1}{1\!+\!\phi^2}\, , \, \eta \! = \!
\frac{\phi}{1\!+\!\phi^2} \, , \, \xi \! = \! \frac{\phi^2}{1\!+\!\phi^2} \, ,
\, \phi \! = \! \frac{\phi_B}{1\!-\!2i\delta/\Gamma},
\end{equation}
and are real on resonance ($\delta = 0$). From the polarizability we then
get the  single-atom differential cross-section:
\begin{equation}
\label{secdiff}
I(\mathbf{k}\boldsymbol{\epsilon}\rightarrow \mathbf{k}'\boldsymbol{\epsilon}')
= \frac{k^4}{16\pi^2}
\left|\bar{\boldsymbol{\epsilon}}'\underline{\alpha}(\mathbf{B})\boldsymbol{
\epsilon}\right|^2 =\frac{3\sigma_0}{8\pi}
\left|\bar{\boldsymbol{\epsilon}}'\mathcal{T}\boldsymbol{\epsilon}\right|^2\\
\end{equation}
where $\sigma_0 = |\alpha_0|^2  k^4/6\pi$ is the total scattering cross-section
at zero magnetic field. As immediately seen, the single atom differential
cross-section~\eqref{secdiff} only
depends on the incoming and outgoing polarizations and is thus completely
insensitive to the change $\mathbf{k}' \to -\mathbf{k}'$. For an isolated atom
there is thus no possible asymmetry when reversing the direction of
observation~\cite{europhys45LT,josaa15LTRS}. As a consequence, the single
scattering signal originating from an assembly of such atoms cannot display any
directional asymmetry and the PHE, if any, must come from a multiple scattering
effect.

\begin{figure}[!]
\includegraphics[width=4cm]{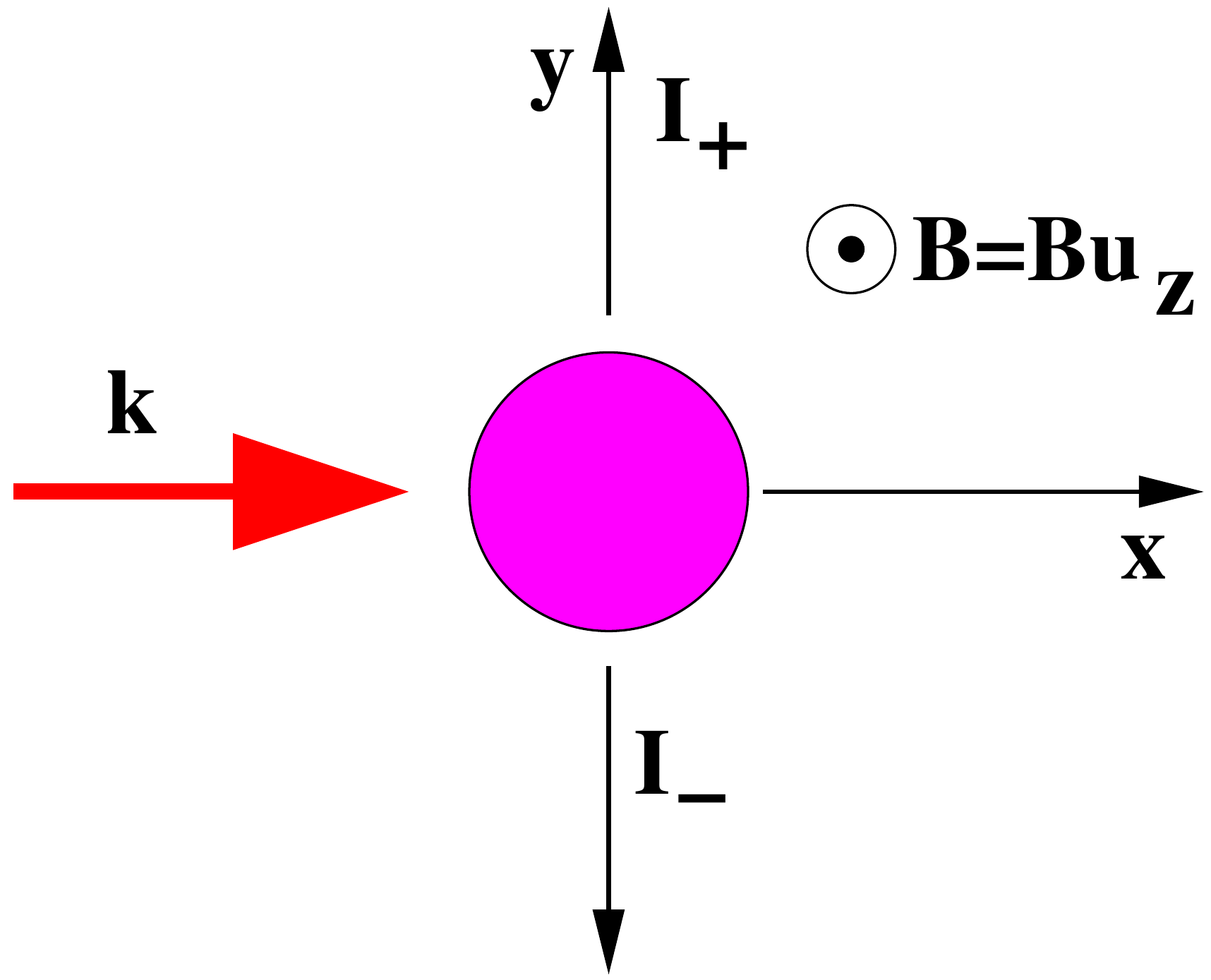}
\caption{\label{hallgeom} (Color online) The photonic Hall geometry. A plane wave
$\mathbf{k}=k\hat{\mathbf{x}}$ is scattered by a cloud of atoms subjected to a
static magnetic field $\mathbf{B} = B \hat{\mathbf{z}}$. 
The Hall current is measured either in the linear polarization channel 
$\hat{\mathbf{y}} \to \hat{\mathbf{x}}$ or in the $\hat{\mathbf{y}} \to \hat{\mathbf{z}}$ one. 
It is defined as
$\Delta I =( I_+ - I_-)$, with $I_\pm$ the configuration-averaged 
differential cross-section along $\pm \mathbf{y}$.}
\end{figure}

\begin{figure}[!]
\includegraphics[width=7cm]{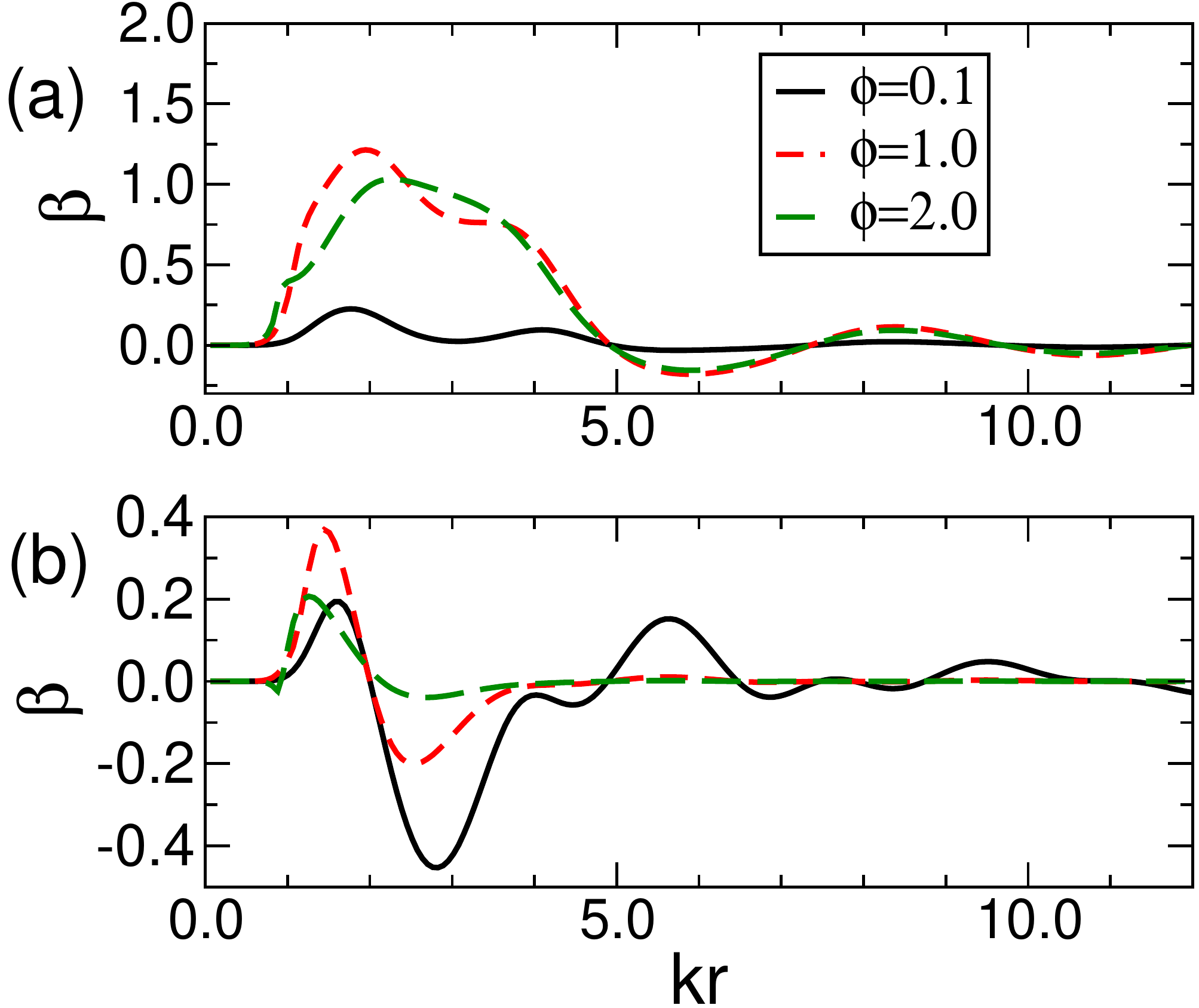}
\caption{\label{2atomnum} (Color online) Two-atom case. Relative Hall current $\beta$ observed
at resonance $\delta =0$ in the linear polarization channels $\hat{\mathbf{y}}
\to \hat{\mathbf{x}}$ (a) and $\hat{\mathbf{y}} \to \hat{\mathbf{z}}$ (b) as a function
of the relative distance $kr$  for
different values of $\phi_B=2\mu B/\Gamma$.}
\end{figure}

Before considering the general case, we first analyze the radiation
properties of two isolated atoms located at positions $\mathbf{r}_1$ and
$\mathbf{r}_2$. This is the simplest possible situation where multiple
scattering plays a role and it will allow us to extract physical insights about
possible mechanisms at work in the PHE. In the Hall geometry (see
Fig.~\ref{hallgeom}), one measures the differential cross sections
$I_{\pm}$ along the up and down directions  $\pm \hat{\mathbf{y}}$ perpendicular both
to the incident light direction $\mathbf{k} = k \hat{\mathbf{x}}$ and to the
magnetic field $\mathbf{B} = B  \hat{\mathbf{z}}$. The total radiation field is
the sum of the fields radiated by the atomic dipoles induced by the incoming and
scattered fields at their respective positions. The exact solution thus involves
the inversion of a linear system of 2 coupled vectorial equations where the
polarizability tensor plays a key role. For a fixed relative distance $r =
|\mathbf{r}_1-\mathbf{r}_2|$ between the atoms, we compute exactly the differential 
cross sections $I_\pm$ and average them
over all possible relative orientations of the atoms to get $\langle I_\pm \rangle$.   
We then extract the
Hall current $\Delta I =\langle I_+ \rangle - \langle I_- \rangle$, 
the mean intensity $I=(\langle I_+ \rangle + \langle I_- \rangle)/2$ and 
the relative Hall 
current $\beta = \Delta I /I$. There are 4 possible
linear polarization channels $\boldsymbol{\epsilon} \to \boldsymbol{\epsilon}'$
for the data analysis. However we first note (and we have numerically checked)
that the Hall currents {\it must be the same} in the linear channels
$\hat{\mathbf{y}} \rightarrow \hat{\mathbf{z}}$ and $\hat{\mathbf{z}}
\rightarrow
\hat{\mathbf{x}}$ since they are related by time-reversal symmetry. Second, 
as we also checked
numerically, the Hall current {\it must  vanish} in the $\hat{\mathbf{z}}
\parallel \hat{\mathbf{z}}$ channel since the polarizations being along the
magnetic field, they are insensitive to the Hanle effect. We are thus left with
the two $lin \perp lin$ channels $\hat{\mathbf{y}} \to \hat{\mathbf{x}}$ and
$\hat{\mathbf{y}} \to \hat{\mathbf{z}}$. Fig.~\ref{2atomnum} summarizes our
numerical results. We see that the Hall current vanishes as $kr \to 0$. Indeed
for very small distances, the two radiating dipoles are always in phase and thus
add up constructively meaning that the two atoms behave like a single scatterer
with a dipole moment twice larger, a situation for which we already know there
is no possible directional asymmetry. We also see that for low $\phi$ values, 
the relative Hall current  decreases in the $\hat{\mathbf{y}} \to
\hat{\mathbf{z}}$ channel, whereas, in the $\hat{\mathbf{y}} \to
\hat{\mathbf{x}}$ channel, it is comparable to the one for high $\phi$ values.
The reason is that, in the $\hat{\mathbf{y}} \to \hat{\mathbf{x}}$ channel, the single scattering 
due to the Hanle effect increases with the magnetic field, such that the background intensity in this
channel gets more and more contaminated (and even dominated at large distances) 
by the single scattering signal. In the  $\hat{\mathbf{y}} \to \hat{\mathbf{z}}$ channel, 
it is always filtered out. 

\begin{figure}[!]
\includegraphics[width=4cm]{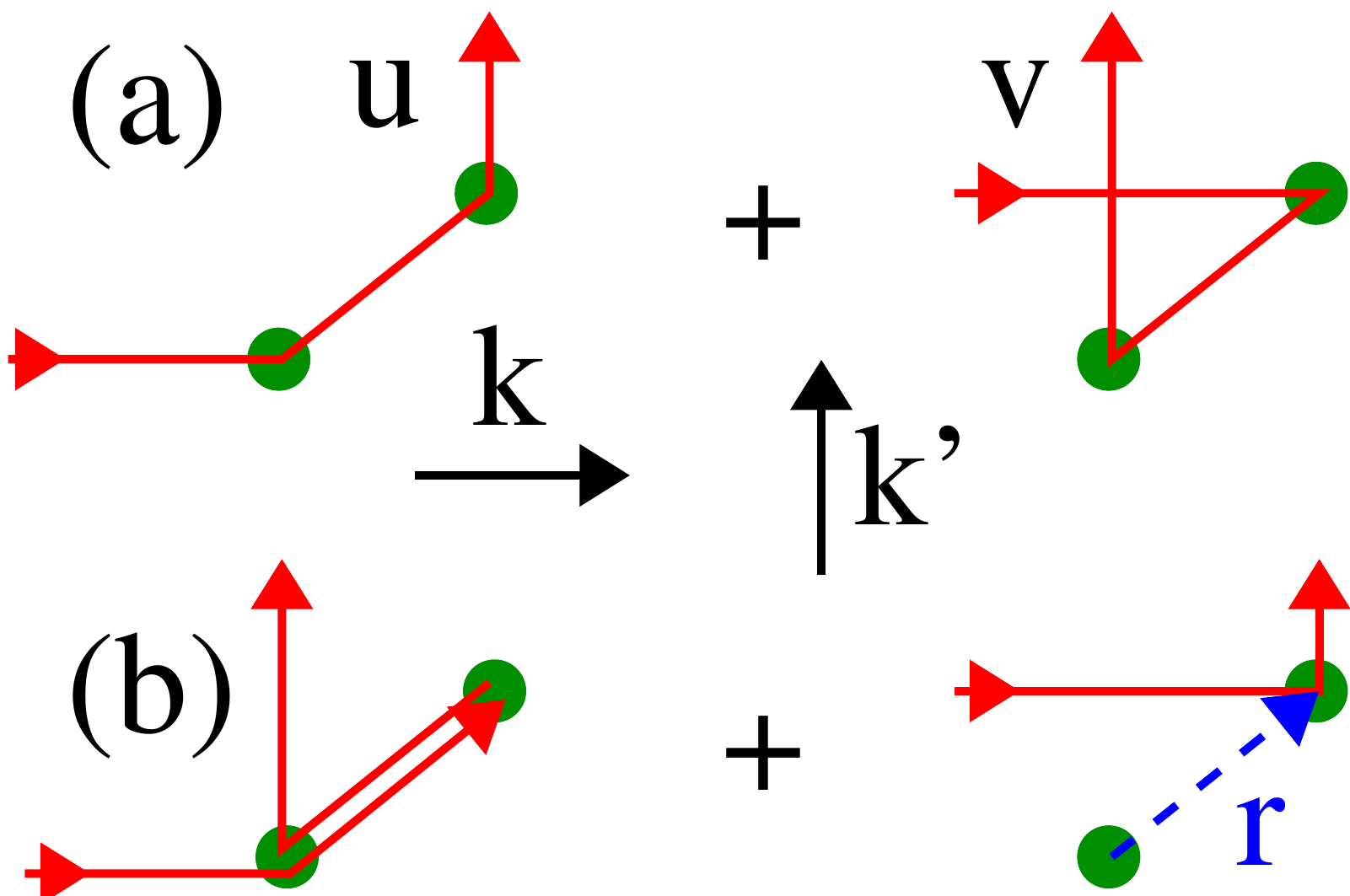}
\caption{\label{2atomdiag} (Color online) Two-atom case. For large distances, the PHE in the
$\hat{\mathbf{y}} \to \hat{\mathbf{z}}$ channel results from the interference
between the two diagrams (a). In the 
$\hat{\mathbf{y}} \to \hat{\mathbf{x}}$ channel an additional contribution comes
from the interference between the two diagrams accounting for 
recurrent scattering (b) (and also between the ones obtained by exchanging the two atoms). }
\end{figure}

To get some insights about the physical processes at work, we consider the case
of a ``dilute" medium $kr \gg 1$ and we expand the field radiated by the two
atoms in powers of $(kr)^{-1}$. Skipping tedious details, 
at leading order the scattered amplitude is obtained from the sum of the 
two diagrams shown in Fig.~\ref{2atomdiag}a, the respective amplitude 
being denoted by $u$ and $v$. Each differential cross-section $I_\pm$ 
contains thus interference terms (i.e. $u\bar{v}+v\bar{u}$) and background 
terms (i.e. $|u|^2+|v|^2$), but since the later are independent of the 
scattering direction, they cancel out in the difference. The Hall current 
is thus solely given by a difference of interference terms, which are 
precisely the crossed terms at the heart of the coherent backscattering effect~\cite{cbs}. 
More precisely, the Hall current reads
\begin{equation}
\label{crossed}
\Delta I = \bigl\langle\delta I(\mathbf{r},\mathbf{B})
\bigl\{\cos{[(\mathbf{k}+\mathbf{k}')\cdot\mathbf{r}]}-
\cos{[(\mathbf{k}-\mathbf{k}')\cdot\mathbf{r}]}\bigr\}\bigr\rangle,
\end{equation}
where $\delta I(\mathbf{r},\mathbf{B})\propto\left| \bar{\boldsymbol{\epsilon}'}\mathcal{T}(\phi)
\boldsymbol{\Delta}_{\mathbf{r}} \mathcal{T}(\phi)
\boldsymbol{\epsilon} \right|^2$, $\langle ... \rangle$ denotes the average over 
the relative orientation of the two atoms and $\boldsymbol{\Delta}_{\mathbf{r}}$ 
is the projector onto the
plane perpendicular to $\mathbf{r}$. After elementary manipulations, 
equation~\eqref{crossed} can be conveniently rewritten as follows:
\begin{equation}
\Delta I =  \left\langle \, \bigl(\delta I(\mathbf{r},\mathbf{B}) - \delta I(\mathbf{r},-\mathbf{B})\bigr)\,
\cos[(\mathbf{k}+\mathbf{k}')\cdot\mathbf{r}]\right\rangle.
\label{asym}
\end{equation}
The Hall current can then be understood as a difference between the two  
configuration-averaged interference effects generated in the \textit{same} 
direction $\mathbf{k}'$, but for opposite directions of the magnetic field. 
The imbalance results from  the interplay between the dipole rotation 
induced by the Hanle 
effect and the transverse projector $\boldsymbol{\Delta}_{\mathbf{r}}$.  
Moreover, since the cosine term depicts very fast oscillations, a 
stationary phase approximation shows that the main contribution 
in Eq.~\eqref{asym} comes from the configurations where the two atoms 
are aligned along the $\mathbf{k}+\mathbf{k}'$ direction. 

Performing the angular average, one gets the Hall current in the
$\hat{\mathbf{y}} \rightarrow \hat{\mathbf{z}}$ channel:
\begin{equation}
\label{hallasympyz}
\Delta I \approx \frac{81}{8k^2} \,
\frac{\phi_B}{|(1-2i\delta/\Gamma)^2+\phi_B^2|^2} \, \frac{\cos{(\sqrt{2}kr)}}{(kr)^4},
\end{equation}
the $\sqrt{2}kr$ term simply corresponding to the value of 
$(\mathbf{k}+\mathbf{k}')\cdot\mathbf{r}$ when these two vectors are parallel to each other. 

In the $\hat{\mathbf{y}} \to \hat{\mathbf{x}}$ channel, a closer inspection 
shows that, at same order, there is an additional
contribution due to diagrams accounting for the interference between recurrent
and single scattering processes (see Fig.~\ref{2atomdiag}b). The final
expression for the Hall current in this channel is quite tedious but simplifies
at resonance $\delta =0$:
\begin{equation}
\label{hallasympyx}
\Delta I \approx \frac{81\sqrt{2}}{8k^2} \, \frac{\phi_B}{(1+\phi_B^2)^3} \,
\frac{\sin{(\sqrt{2}kr)}}{(kr)^3} \left(1+\cos{(2kr)}\right).
\end{equation}

\begin{figure}[!]
\centerline{\includegraphics[width=7cm]{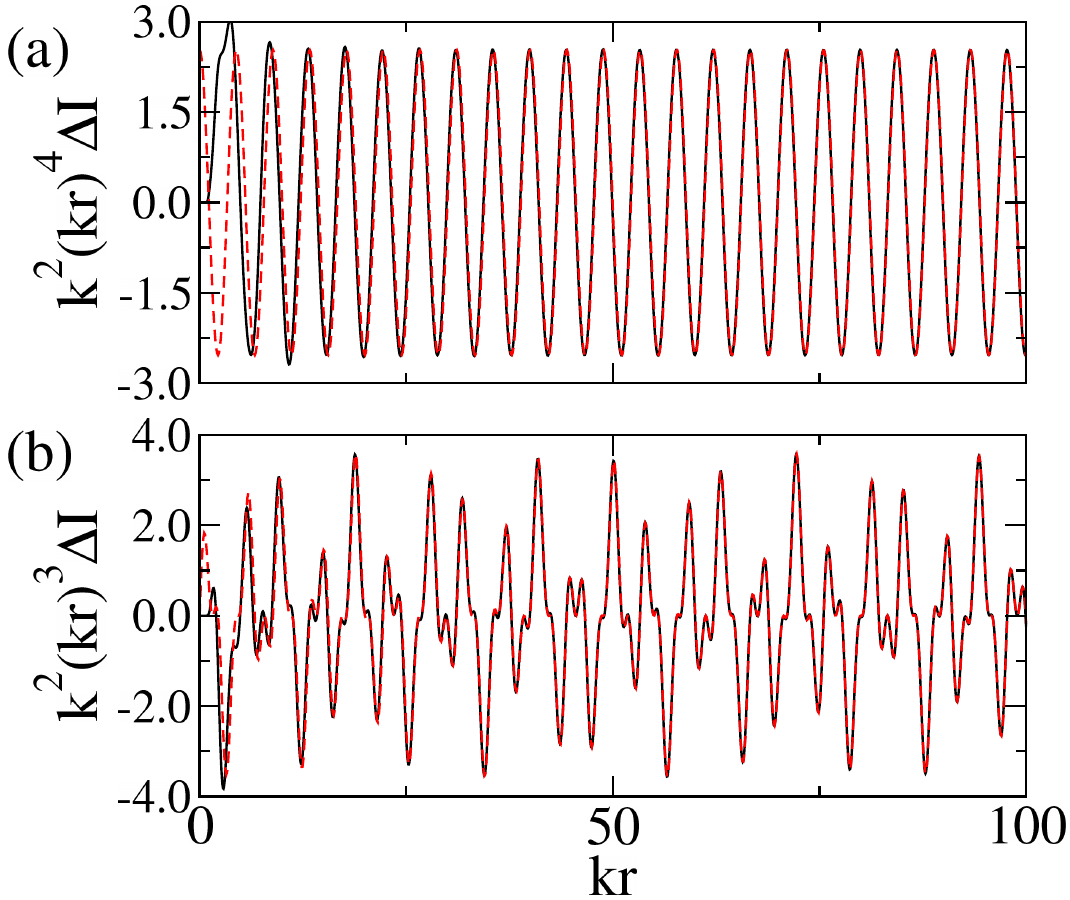}}
\caption{\label{kr_large} (Color online). Normalized Hall current in the
two-atom case. Comparison between the numerical calculations (plain curves) and
the analytical predictions Eqs.~\eqref{hallasympyz},\eqref{hallasympyx} (dashed
curves) as a function of $kr$ for $\delta=0$ and
$\phi_B=1$.  (a) $\hat{\mathbf{y}} \rightarrow \hat{\mathbf{z}}$ channel. (b)
$\hat{\mathbf{y}} \rightarrow \hat{\mathbf{x}}$ channel. The excellent agreement
emphasizes that the underlying physics of the PHE is fully catched by the diagrams depicted in Fig.~\ref{2atomdiag}.}
\end{figure}

As shown in Fig.~\ref{kr_large} for $\delta=0$ and $\phi_B=1$, the agreement
between the numerical computations and these asymptotic results proves excellent
as soon as $kr \gtrsim 10$. In particular it is quite remarkable
that recurrent scattering is essential to reproduce the additional oscillations
observed in the $\hat{\mathbf{y}} \rightarrow \hat{\mathbf{x}}$ channel and due
to the $\cos 2kr$ term in Eq.~\eqref{hallasympyx}.
We have also numerically checked that the agreement is still excellent in both
channels when $\delta \not= 0$. This gives clear evidence that the physical
mechanism at the heart of the PHE for atomic scatterers is the interference
between the scattering processes depicted by the diagrams in
Fig.~\ref{2atomdiag}. In addition, the fact that, in the small $\phi_B$ limit, the Hall current is
proportional to $\phi_B$ (i.e. to the atomic dipole rotation) actually
reflects that  only one active scatterer would be enough to
generate a Hall current. Finally, one must also mention that, contrary to the case studied
in Ref.~\cite{prl75vT},  both our analytical and numerical
results show that there is still a PHE when the antisymmetric part of the self-energy 
($\Sigma\propto\underline{\boldsymbol{\alpha}}(\mathbf{B})$) 
has a vanishing real part (i.e. at resonance $\delta=0$), the exact dependence on $\delta$ being a smooth
bell-shaped curve.

Assuming an isotropic propagation in the medium (with a
mean free path $\ell$), one can further average the preceding expressions over
all possible positions of the two scatterers. As seen from
Eqs.~\eqref{hallasympyz} and \eqref{hallasympyx}, the interference effect leads
to fast
decreasing oscillations at the wavelength scale. Since $\lambda \ll \ell$, the
main contribution to the Hall
effect will arise for scatterers separated by $r \sim \lambda$. The probability
of finding two such scatterers scales with the optical density $\rho =
n\lambda^3$, $n$ being the scatterer density; we thus expect the actual relative Hall current $\beta$ to be
smaller by a factor
$(k\ell)^{-1}$ than the average background~\cite{propagation}. Obviously, the exact result will
depend on the geometry of the medium and on the optical
thickness. Nevertheless, the PHE in a assembly of resonant point scatterers,
even if rather small, should be measurable.

\begin{table}[!]
\caption{\label{brutenum}  The left side displays the relative
Hall currents $\beta$ at $\delta = 0$ and $\phi_B = 1$, in the Hall geometry, for $500$ atoms uniformly
distributed inside a sphere with optical density $\rho \approx 0.65$. 
The number of disorder configurations is $6\times 10^5$. The expected 
accuracy of the order of $10^{-3}$ is confirmed when comparing to the
relative currents $\beta$ (see right side) in the  
$\mathbf{k} \parallel \mathbf{B}$ configuration, when no PHE should show up.
}
\begin{ruledtabular}
\begin{tabular}{cdcd}
\multicolumn{2}{c}{$\mathbf{k} \perp \mathbf{B}$} &
\multicolumn{2}{c}{$\mathbf{k} \| \mathbf{B}$}  \\
\hline
$\hat{\mathbf{z}} \rightarrow \hat{\mathbf{x}}$ &  0.0579 & 
$\hat{\mathbf{x}} \rightarrow \hat{\mathbf{z}}$ &  0.00267 \\
$\hat{\mathbf{y}} \rightarrow \hat{\mathbf{z}}$ &  0.0545 & 
$\hat{\mathbf{x}} \rightarrow \hat{\mathbf{x}}$ & -0.00014 \\
$\hat{\mathbf{y}} \rightarrow \hat{\mathbf{x}}$ & -0.0088 & 
$\hat{\mathbf{y}} \rightarrow \hat{\mathbf{z}}$ & -0.00121 \\
$\hat{\mathbf{z}} \rightarrow \hat{\mathbf{z}}$ &  0.0022 & 
$\hat{\mathbf{y}} \rightarrow \hat{\mathbf{x}}$ & -0.00004\\
\end{tabular}
\end{ruledtabular}

\end{table}

To bring a numerical confirmation of our findings, we have considered $N=500$
atoms uniformly
distributed inside a sphere at an optical density $\rho \approx 0.65$ 
and illuminated by a plane wave set on resonance. This leads to 
$k\ell=4\pi^2/3\rho \approx 20$ at $B=0$. Such a value is
difficult to achieve in a real experiment but has already been
obtained~\cite{strontium}. The optical thickness along a diameter of the sphere is $b \approx
3.5$.  The magnetic field value
has been set at $\phi_B=1$. To obtain the total radiated field, we have 
solved the corresponding system of $3N$ linear equations~\cite{pinheiro} and we have computed the
various quantities of interest averaged over $6\times 10^5$ configurations, 
such that we expect an accuracy of the order of $10^{-3}$.    
To stress the existence of the PHE, we have also computed $\beta$ in
the geometry $\mathbf{k}\parallel \mathbf{B}$ where no
PHE should be observed; as depicted by the right side of table~\ref{brutenum}, 
the corresponding values are effectively at most of the order of few  $10^{-3}$. 
Up to this accuracy, the numerical results in the geometry 
$\mathbf{k} \perp \mathbf{B}$, are in a qualitative agreement 
with the two-atom case: there is no Hall current  in the $\hat{\mathbf{z}}
\rightarrow \hat{\mathbf{z}}$ channel. It is about the same
order of magnitude in the two
conjugate channels $\hat{\mathbf{y}} \rightarrow \hat{\mathbf{z}}$  and
$\hat{\mathbf{z}} \rightarrow \hat{\mathbf{x}}$ ($\beta \approx 5.5\%$). 
Finally, it is larger in
these channels than in the $\hat{\mathbf{y}} \rightarrow \hat{\mathbf{x}}$
channel ($|\beta| \approx 1\%$). To enforce the validity of our mesoscopic description, 
we have computed $\beta$ for different values of $k\ell$, 
while keeping fixed the optical thickness $b$. The results are displayed 
in table~\ref{scaling} and, as expected, the product 
$\beta k\ell$ is, within the statistical errors, independent of $k\ell$. 
Finally, one must note that the values found here are in the percent range and are much
larger (by three orders of magnitude) than observed with classical scatterers
\cite{nature381RT} although the magnetic field is smaller by two or three orders
of magnitude. This arises because the Zeeman effect in highly-resonant 
atomic scatterers induces a "giant" dipole rotation  which amplifies the PHE. 

\begin{table}[!]
\caption{\label{scaling}  Relative Hall current $\beta$ in the 
$\hat{\mathbf{z}} \rightarrow \hat{\mathbf{x}}$ and 
$\hat{\mathbf{y}} \rightarrow \hat{\mathbf{z}}$ channels at a fixed 
optical thickness $b=1.3$. Within the numerical accuracy, the results 
indicate that $\beta$ scales like $1/(k\ell)$.}
\begin{ruledtabular}
\begin{tabular}{cccc}
$k\ell$ & 40.00 &  63.25 & 89.44 \\ 
$\beta \times k\ell$ & 4.0(1) &  3.8(3) &  4.0(3) 
\end{tabular}
\end{ruledtabular}
\end{table}

In summary, on the basis of numerical and analytical calculation, we
have qualitatively and quantitatively explained the underlying
interference effect at the origin of the
photonic Hall effect for resonant point-dipole scatterers. Our
results show that the effect, albeit small,  should be observable in cold
atomic vapors. Possible extensions of this work would consist first in
developing the diagrammatic analysis to an arbitrary number of scattering
events and, second, in accounting for internal degeneracies in the atomic
groundstate. This would give a quantitative and comprehensive description of
the photonic Hall effect in cold atomic clouds.

The Authors thank G.L.J.A.~Rikken and B.A.~Van~Tiggelen for fruitful
discussions.


\begin{thebibliography}{99}
\bibitem{propagation} Since propagation through the effective medium is affected
by magneto-optical effects, the antisymmetric part of the refractive index also
contributes to the Hall current but with a scaling factor $(k\ell)^{-2}$.
\end{thebibliography}
\end{document}